# Long-term and Real-time High-speed Underwater Wireless Optical Communications in Deep Sea

Jialiang Zhang, Sujing Wang, Ziqi Ma, Guanjun Gao, Yonggang Guo, Fei Zhang, Shanguo Huang and Jie Zhang

*Abstract*—Seafloor observation network can perform all-weather, long-term, continuous, real-time, and in-situ observation of the ocean by combing various observation methods including cabled seafloor nodes, self-contained nodes, as well as mobile platforms, where reliable and long-term high-speed underwater wireless communication becomes an essential demand. Recently, underwater wireless optical communication (UWOC) has emerged as a highly promising solution and is rapidly becoming a research hotspot for meeting this requirement. In this article, we demonstrate the experiment and application of high-speed UWOC system for deep sea seafloor observation network. To the best of our knowledge this is the first long-term real-time deep-sea UWOC link with bitrate as high as 125 Mbps. Between 30 m distance and at a depth of 1650 m, two-way Ethernet UWOC links are realized with 125 Mbps direction-adjustable green light link and 6.25 Mbps non-line-of-sight (NLOS) blue light link. High quality video transmission of 8K 30 FPS and 4K 120 FPS are realized through high-speed 125 Mbps green light link, with 100% peak signal-to-noise ratio (PSNR) agreement, showing the capability of transmitting high-quality videos lossless. The 30-day long-term measurement results show that the BER performance of both 125 Mbps and 6.25 Mbps links is lower than $10^{-5}$, proving the stability and reliability of this UWOC system at depth of 1650 m. By measuring the light attenuation of deep-sea water sample, the maximum transmission distance for the green and blue light links are estimated to be 117.7 and 128.3 m with considering the geometry loss, which can be extended to 231.6 and 337.5 m without geometry loss. As the first long-term and real-time UWOC system in deep sea, we believe this demonstration can provide valuable experience for further UWOC studies and converged ocean observation networking with cabled and cable-less observation platforms.

" This work was supported in part by National Key Research and Development Program of China under Grant 2022YFB2903303, National Natural Science Foundation of China under Grant 62371064 and U22A2012, and BUPT Excellent Ph.D. Students Foundation under Grant CX2022213" (Corresponding author: Guanjun Gao and Yonggang Guo). Jialiang Zhang and Sujing Wang are co-first authors.

Jialiang Zhang, Ziqi Ma, Guanjun Gao, Shanguo Huang and Jie Zhang are with the State Key Laboratory of Information Photonics and Optical Communications, Beijing University of Posts and Telecommunications, Beijing 100876, China (e-mail: jialiangzhang@bupt.edu.cn, MZQ2021@bupt.edu.cn, ggj@bupt.edu.cn, shghuang@bupt.edu.cn and lgr24@bupt.edu.cn).

Sujing Wang, Yonggang Guo and Fei Zhang are with the State Key Laboratory of Acoustic, The Institute of Acoustics of the Chinese Academy of Sciences, Beijing 100190, China (e-mail: wangsujing@mail.ioa.ac.cn, guoyg@mail.ioa.ac.cn, and zhangfei@mail.ioa.ac.cn).

## I. INTRODUCTION

CABLED seafloor observation network is a new platform for human to observe and comprehend ocean, which uses undersea fiber cable to establish power and information connection between shore station and various seafloor observation nodes, enabling all-weather, long-term, continuous, real-time, high-resolution, high-precision and in-situ observation of the ocean from the sea floor to the sea surface [1]. Despite the cabled connections provide abundant feeding power and transmission bandwidth, the cover range and flexibility of the cabled seafloor observation nodes are also limited. Once the seafloor observation nodes are deployed and connected, it is challenging to change their positions or reconnect the cables.

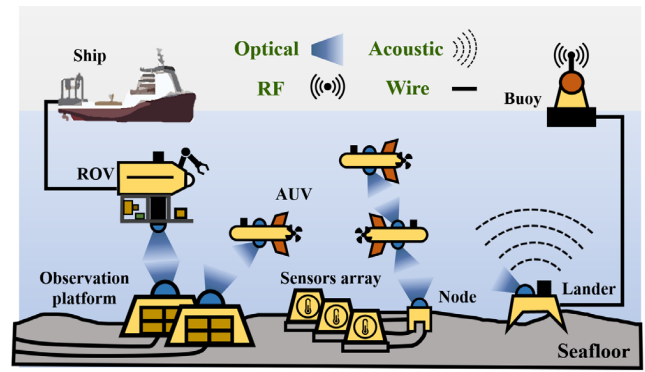

**Fig. 1.** The concept of cabled and cable-less converged observation network.

Cable-less observation method provide a complementary observation approach [2]. For seafloor observation network, the combination of cabled and cable-less methods can improve its coverage area and flexibility, the concept of which is shown in Fig. 1. The cable-less observation platforms include self-contained in-situ observation node, such as bottom landers, and surface buoys, and mobile platforms such as autonomous underwater vehicles (AUVs), remotely operated vehicles (ROVs), et. al. For realizing such a converged seafloor observation networking, it is an essential demand to provide high-speed and reliable real-time wireless communication between the cabled and cable-less observation platforms.

Underwater wireless optical communication (UWOC) is attracting interest rapidly and becomes a competitive method for information connection between cabled and cable-less platforms, due to its advantage of lower power consumption, and higher bandwidth compared with acoustic and radio frequency (RF) methods [3-4]. Recently, notable progress and advances on UWOC have been achieved, focusing on channel modelling, advanced processing algorithm, system bitrate and transmission distance improvement [5-8]. For underwater channel modelling, a weighted Gamma function polynomial (WGFP) is proposed to model the impulse response of general UWOC MIMO links with arbitrary numbers of light sources and detectors [5]. Under laboratory environment, a 500 Mbps UWOC offline experiment has been achieved across 100 m distance utilizing a green laser diode (LD) with tiny transceiver aperture and precise light alignment [6]. Moreover, 500 Mbps transmission over 200 m distance has also been reported in a swimming pool with advanced offline equalization algorithm, using blue laser diode [7]. Recently, a real-time UWOC protype is realized with light-emitting diode (LED), which transmits at a rate of 80 Mbps over 32 m [8].

Several sea-trial UWOC experiments have also been demonstrated, providing valuable experiences and guidance. For instances, C. Pontbriand et al. from Woods Hole Oceanographic Institution (WHOI) realize 5 Mbps transmission over 200 m deep clear water at Bermuda in April, 2008 [9]. The same research group further achieve wireless data harvesting from a subsea node enabled by UWOC devices with bandwidth up to 10 Mbps [10]. In 2018, G. Cossu et al. demonstrate an experiment of real-time 10 Mbps UWOC transmission over 10 m sea water between a fixed and a mobile node [11]. T. Sava et al. from Japan Agency for Marine-Earth Science and Technology (JAMSTEC) demonstrate a 20 Mbps UWOC system over 120 m at depth of 700 m between two underwater vehicles [12]. In October 2022, S. Ishibashi et al. from JAMSTEC report their latest sea-trial experiment using multi-beam and multi-PMT based transceivers, successfully demonstrating the feasibility of demodulating 1 Gbps data over 100 m at depth of 800 m [13].

However, for most of these UWOC sea-trial demonstrations, the real-time communication bitrate is limited to only 20 Mbps, and the communication links are temporarily constructed, due to the difficulty of remote power supply, cabled system construction and challenge of UWOC transceiver alignment in deep-sea environment. And there are also challenges for non-deep-sea environment, including interference of ambient light source, poor water quality, and difficulty in alignment between mobile platforms, limiting the data rate and transmission distance of UWOC systems. Without long-term and real-time verification, the feasibility and reliability of UWOC system cannot be sufficiently proved for deep-sea observation applications between cabled and cable-less nodes, which is of extremely difficult and great value.

In this article, we demonstrate long-term and real-time UWOC links on deep-sea seafloor observation network experiment system of China, where the long-term power supply and Ethernet connection is provided for UWOC systems, enabling online performance monitoring and control. The UWOC links achieve two-way real-time Ethernet connections between two deep-sea seafloor observation nodes located at a distance of 30 m and at a depth of 1650 m, realizing 125 Mbps direction-adjustable green light link and 6.25 Mbps non-line-of-sight (NLOS) blue light link. To the best of our knowledge, this is the first time that long-term and real-time UWOC links are constructed in deep-sea environment, with wireless communication bitrate as high as 125 Mbps. The 30-day measurement results show that the bit-error-rate (BER) of both 125 Mbps and 6.25 Mbps links are lower than $10^{-5}$, below the forward error correction (FEC) threshold of our UWOC transceivers, proving its stability and reliability. By measuring the light attenuation of deep-sea water sample, the maximum transmission distance for green and blue light links are estimated to be 117.7 m and 128.3 m considering the geometry loss, which can be extended to 231.6 m and 337.5 m if geometry loss is excluded. As the first long-term and real-time UWOC system in deep sea, the feasibility and reliability of cabled and cable-less converged seafloor observation system is verified, which we believe can provide valuable experience for further UWOC studies and ocean observation applications.

## II. SYSTEM DESIGN

Two kinds of real-time transceiver are designed and developed to demonstrate both UWOC links, the structures of the two transceivers are depicted in Fig. 2. The processing module in both transceivers is used for generating the transmission signal and processing the received signal, while the acquisition and control block is utilized for optical power control and collection. Both devices are connected by 100M Ethernet interface, supporting remote control and various communication services using Ethernet. Transceiver A emits 525nm green laser and transceiver B emits 450 nm blue laser. High-sensitivity photomultiplier tube (PMT) with 46 mm aperture is equipped in transceiver A, high-speed PMT with 8 mm aperture is equipped in transceiver B. In front of each PMT, an optical filter is inserted to remove backscattering light for the purpose of duplex communication, and a low-cost liquid crystal (LC) is used as variable optical attenuator.

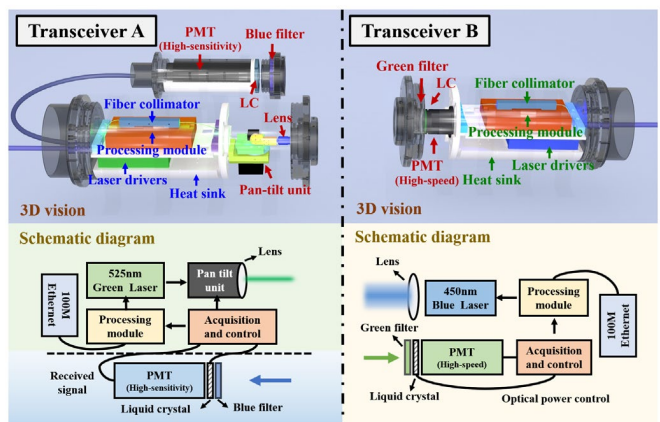

**Fig. 2.** Structures of transceiver A and B in three-dimension vision and schematic diagram.

The transceiver A has two cabins, the main cabin has a pan-tilt unit (PTU), it is a two-axis gimbal with two servos. The PTU module can be manually controlled online to adjust the optical path for the alignment from A to B. There is a high-sensitivity PMT in the auxiliary cabin for transceiver A, while the PMT and other modules are all integrated into a single cabin for transceiver B. On-board signal processing, optical power acquisition and receiver control are performed, where the gain of PMT and attenuation of LC can be flexibility adjusted according to the optical power acquisition result [14].

The optical link path between A and B, and the two transceivers are equipped on the ocean bottom node (OBN) platforms. The green signal light transmits data from A to B at a rate of 125Mbps. On the other hand, the blue signal light transmits data from B to A at 6.25Mbps. It is important to note that the blue signal light works as a non-line-of-sight (NLOS) link, where the received light is primarily reflected or scattered.

To ensure a stable underwater wireless optical communication (UWOC) link in the complex and dynamic undersea environment, optical power monitor and control at the receiver is an effective approach. The signal amplitude is measured in voltage units, subsequently captured and collected by the acquisition and control module's analog-to-digital converter (ADC). The control voltage of the liquid crystal and the gain of the PMT determine the signal amplitude, which need to be maintained within a specified range for optimal communication performance. Prior to the sea-trial experiment, calibration of the function is performed to establish the relationship between these three parameters. Based on the calibration results, the input optical power and PMT gain can be adjusted to obtain an appropriate signal amplitude during the practical undersea usage, resulting in an optimal receiving condition.

The transmitted optical power is 2.36W for green laser and 2.1W for blue laser, and laser beam divergence is 0.53 ° for green light and 3.83 ° for blue light, both of which can be manually tuned during system integration. The modulation format is On-Off Keying (OOK) for green light link and 4-Pulse-position modulation (4-PPM) for blue light link. The FEC used is concatenated BCH (2040, 1930) and BCH (3860, 3824) with 7% redundancy. The total link loss budget is 82.77 dB and 100.54 dB for green light link and blue light link, respectively. Before sea trial these parameters are measured and verified by authorized certification testing organization, China Academy of Information and Communications Technology (CAICT).

The UWOC system is sealed within titanium alloy cabins and has successfully passed a series of environmental tests, including vibration, temperature cycling, and water pressure resistance tests. It is capable to withstand water pressure at depths of up to 3000 meters. The titanium alloy cabins ensure long-term resistance to corrosion and water pressure, and is designed to be isolated from frameworks of observation nodes by insulation pads. The framework of observation nodes is equipped with sacrificial anodes to avoid corrosion. Through the above design and environmental tests, the system is capable of long-term resistance of seawater corrosion and pressure in deep sea.

## III. DEEP-SEA DEPLOYMENT

The deployment was conducted in Oct. 24, 2022. Transceivers A and B are mounted on two OBN platforms of a seafloor scientific observation network, connected with its undersea cable landing station (CLS). The power and Ethernet of OBN is provided and connected with underwater primary node (PN), which performs voltage conversion from high to low, and converges the Ethernet tributaries to 100G wavelength. Through such a deployment, long-term continuously UWOC performance measurement can be performed, including remote Ethernet bandwidth test, video transmission test and remote control.

The distance between the two OBN platforms is 30 m, and there is also 45° angle offset due to the restriction and difficulties of deep-sea deployment. Under such a large angle deviation and deep-sea condition, both the blue and green light UWOC link cannot be easily aligned. At the transmitter side, the beam direction of the green light is remotely controlled by a built-in PTU to realize adequate alignment, as shown in Fig. 3. It is estimated that the light spot diameter arriving at detector is around 0.55 m, which covers and exceeds the detector's detection area considerably under the aligned condition. During the deployment, the cases of misaligned and aligned for green light link has been captured by ROV.

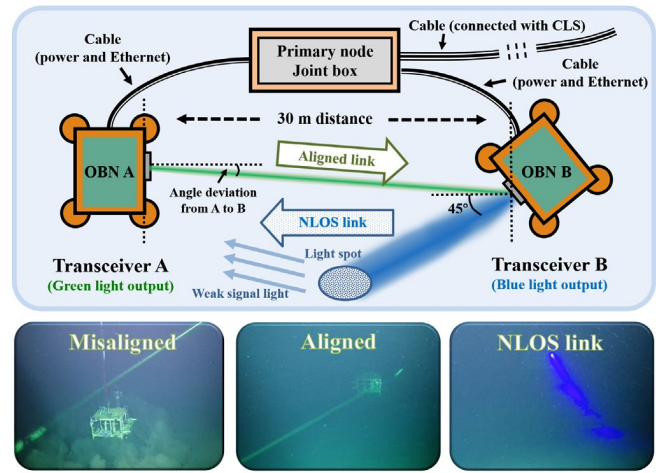

**Fig. 3.** Experimental setup of UWOC system deployed in seafloor scientific observation network.

On the other hand, for the purpose of compact integration, there is no PTU for blue light link in transceiver B. And the blue light link is found to be fully NLOS under such a large angle offset, where the light beam is apparently deviated and hit the seabed, observed from the camera of ROV, also shown in Fig. 3. Under such condition, the blue light signal received at transceiver A is the reflected light from the seabed and the scattering light propagated by suspended particles.

## IV. RESULTS

After the deep-sea deployment, the basic function checks and performance measurements of the UWOC links are performed. Firstly, part of the Ethernet connection performance results is analyzed as shown Fig. 4, which are recorded through the Ethernet analyzer in Oct. 27, 2022. The effective Ethernet bandwidth carried by green light link and blue light link is 97.4 Mbps and 5.5 Mbps, respectively, due to the constraint of the 100M Ethernet interface.

During the test, the occurrence of bit errors before FEC is also monitored. It is observed that there is no packet loss on the green light link. Moreover, the bit errors per second (BEPS) are less than 4 for green light link, which can be easily corrected after FEC. However, for the blue light link under NLOS condition, there are two packet loss events, which occurs when the BEPS is 63 and 85 respectively, making 12 Ethernet packet loss in total. Compared to the 125 Mbps green light link, the link loss budget of 6.25 Mbps blue light link is much higher due to its higher sensitive modulation of 4-PPM and larger detection aperture. However, as an NLOS link, the received blue light signal is very weak after reflecting from seabed and suspended particle scattering. Under the influence of sea current and sea lives, the signal intensity becomes unstable, resulting in packet losses.

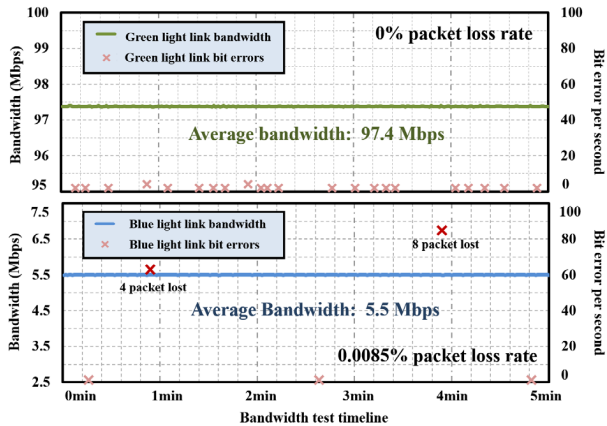

**Fig. 4.** The Ethernet bandwidth performance for both links.

After some receiver parameters optimization such as PMT gain and LC attenuation, we start a long-term UWOC link performance measurement. A 30-day BER monitoring test is conducted, which runs continuously for one hour each day. As depicted in Fig. 5, the results indicate that for both the 125 Mbps green light link and 6.25 Mbps blue light link, the 30-day BER results before FEC are below $10^{-5}$ and $10^{-7}$, respectively. The BER performance results for both links are all below the FEC threshold, which proves the reliability of the UWOC system.

Next, high-definition (HD) video transmission through deep-sea UWOC link is also realized in this demonstration. Two HD video formats are tested, including 8K video with 7680×4320 resolution at 30 frames per second (FPS) and 4K video with 4096×1860 resolution at 120 FPS. It is observed that for the 125 Mbps green light link, the maximum bandwidth occupied by 8K and 4K video is 50.5 Mbps and 34.4 Mbps, respectively. Moreover, the peak signal-to-noise ratio (PSNR) of the source and received videos are completely the same for all the frames, indicating no loss in quality during UWOC-based HD video transmission. This demonstration marks the successful real-time transmission of an 8K HD video through a wireless communication method in the deep sea, which is a significant achievement in UWOC studies.

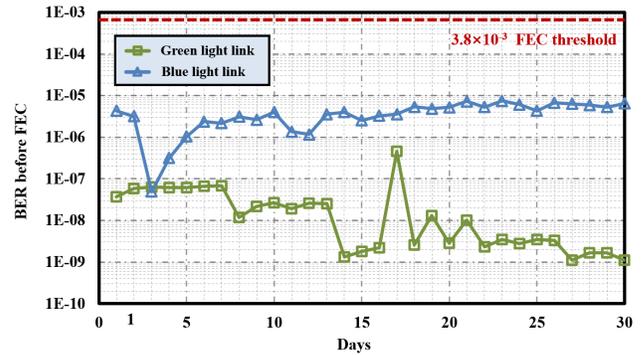

**Fig. 5.** The long-term BER test in 30 days for both links.

## V. ANALYSIS

After the deployment of UWOC system, we also obtain the deep-sea water sample at 1650 m depth, which can be used to further estimate the maximum transmission distance of our UWOC system.

In laser-based UWOC systems, the link loss of is mainly introduced by water attenuation and geometric loss. The water attenuation of light is commonly described using exponential decay model [15], which can be expressed as follows:

$$L_{att} = \exp(-c \cdot Z), \quad (1)$$

where the $c$ is the diffuse attenuation coefficient and $Z$ is the distance of the optical link. According to the measurement of deep-sea water sample, the attenuation coefficient $c$ is 0.069 (0.298 dB/m) for blue light link and 0.082 (0.358 dB/m) for green light link.

In addition to the water attenuation loss, geometric loss of UWOC link also arises if the signal light cannot be completely collected by receiver detector. The geometric loss is inversely proportional to the square of the distance, which can be expressed as follows:

$$L_{geo} = \frac{k}{Z^2}, \quad (2)$$

where $k$ is constant and related to the half-angle beam divergence $\varphi$, and detector aperture. According to our adjustment and measurement before sea-trial deployment, the laser beam divergence is 3.83° for blue light link, and 0.53° for green light link.

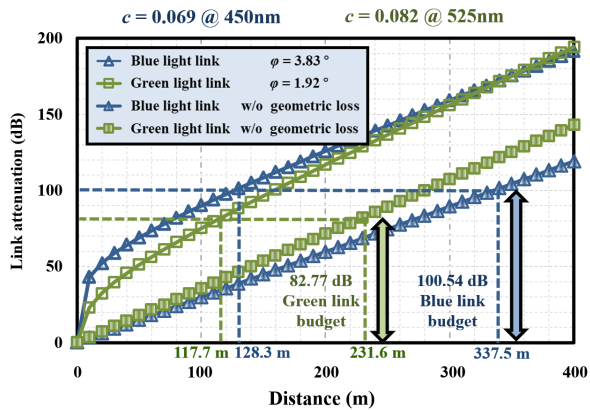

**Fig. 6.** Maximum transmission distance estimation for the two UWOC links.

Assuming a link-aligned condition, the maximum transmission distance for both links are estimated considering with and without geometric loss, as shown in Fig. 6. It is observed that for our current transceiver configuration, the maximum transmission distance is 117.7 m for 125 Mbps green light link and 128.3 m for 6.25 Mbps blue light link, when geometric loss is included. If the receiver detector area is large enough to completely collect the signal light with narrow beam divergence, an ideal receiving condition without geometric loss can be achieved. Under such condition, the maximum transmission distance for 125 Mbps green light link and 6.25 Mbps blue light link can be extended to 231.6 m and 337.5 m, respectively, which we believe is still challenging to approach in practical undersea condition.

## VI. Conclusion

We demonstrate for the first time a long-term, real-time high speed UWOC link in deep sea seafloor scientific observation network with bitrate as high as 125 Mbps. Two-way Ethernet UWOC links over distance of 30 m is realized, including 125 Mbps directional adjustable green light link and 6.25 Mbps NLOS blue light link. The long-term measurement results show that BER of both links are lower than $10^{-5}$, proving its stability and reliability. For the first time, 8K HD video is wirelessly transmitted without quality loss in deep sea, based on the high-speed UWOC link. The maximum transmission distance for green and blue light links are estimated to be 117.7 and 128.3 m, respectively. We expect our effort on deep-sea UWOC system can provide valuable experience for further UWOC studies and converged ocean observation networking between cabled and cable-less platforms.

## Acknowledgment

The authors would like to thank Da Kang and Qi Zheng from The Institute of Acoustics of the Chinese Academy of Sciences for their help on sea-trial deployment in this research.

## Biographies

**Jialiang Zhang** received the B.E. degree from the Beijing University of Posts and Telecommunications (BUPT), Beijing, China, in 2015. He is currently working toward the Ph.D. degree in optical engineering at the Beijing University of Posts and Telecommunications, Beijing, China, BUPT. His current research interests include underwater wireless optical communication.

**Sujing Wang** received the B.S. degree in Electronic Science &Technology from University of Science and Technology of China, Hefei, China, in 2009 and the Ph.D. degree in Earth and Space Exploration Technology from University of Chinese Academy of Sciences, Beijing, China in 2015. He is currently an Associate Research Fellow in Key Laboratory of Acoustics, Institute of Acoustics, Chinese Academy of Sciences. His research interests include seafloor observation network and ocean geophysical exploration.

**Ziqi Ma** received the B.E. degree from the University of Science & Technology Beijing (USTB), Beijing, China, in 2021. He is currently working toward the Ph.D. degree in optical engineering at the Beijing University of Posts and Telecommunications (BUPT), Beijing, China. His current research interests include Underwater Optical Wireless Communication.

**Guanjun Gao** received the B.E. and Ph.D. degrees from the Beijing University of Posts and Telecommunications (BUPT), Beijing, China, in 2006 and 2012, respectively. He is currently an associate professor and the director of optical communication network center in State of Key Laboratory of Information Photonics and Optical Communications, BUPT. His current research interests include underwater wireless optical communication, undersea fiber cable system and networks, and physical-layer aspects of optical communications.

**Yonggang Guo** received the B.E. degree from the Ocean University of Qingdao, Qingdao, China, in 1998, and the Ph.D. degree from the Ocean University of China in 2004. He is currently a professor in Key Laboratory of Acoustics, Institute of Acoustics, Chinese Academy of Sciences. His research interests include seafloor observation network and ocean geophysical exploration.

**Fei Zhang** received the Maser degree from university of science and technology Beijing in 2015. He is currently a senior engineer in Key Laboratory of Acoustics, Institute of Acoustics, Chinese Academy of Sciences. His research interests include seafloor observation network and ocean engineering.

**Shanguo Huang** (Member, IEEE) received the Ph.D. degree from the Beijing University of Posts and Telecommunications (BUPT), Beijing, China, in 2006. He is currently a professor and director of the State Key Laboratory of Information Photonics and Optical Communications. He is also the Dean with the School of Electronics, BUPT. His research interests include optical communication networking.

**Jie Zhang** received the Ph.D. degree in electromagnetic field and microwave technology from Beijing University of Posts and Telecommunications (BUPT), Beijing, China, in 1993 and 1998, respectively. He is currently a professor and the Dean of School of Integrated Circuits, BUPT. He has published more than 300 technical papers, authored 8 books and submitted 17 ITU-T recommendations and 6 IETF drafts. He has served as a TPC member of various conferences such as OFC, ACP, OECC, ONDM, etc. His research interests include secure optical transport networks.